# Universal Dzyaloshinskii–Moriya interaction dictates pairing in unconventional superconductor families

Baishun Yang[1,†], Yida Chu[2,†], Xuelei Sui[2], Haiqing Lin[3,2*], Shijie Hu[2,4*], and Bing Huang[2,4*]

*1. CIC nanoGUNE BRTA, Tolosa Hiribidea 76, 20018 San Sebastián, Spain.*

*2. Beijing Computational Science Research Center, 100193, Beijing, China.*

*3. School of Physics and Institute for Advanced Studies of Physics, Zhejiang University, Hangzhou, 310058, China*

*4. School of Physics and Astronomy, Beijing Normal University, Beijing 100875, China*

*† These authors contributed equally.*

Email: hqlin@zju.edu.cn, shijiehu@csrc.ac.cn, bing.huang@csrc.ac.cn

**The collinear-antiferromagnetic spin-fluctuation paradigm has long guided unconventional superconductivity research, yet fails to reconcile the noncollinear spin phenomena observed across cuprates, iron-based superconductors, and nickelates. Using extensive first-principles calculations and unbiased large-scale DMRG simulations, we show that Dzyaloshinskii–Moriya interaction (DMI)—arising from local inversion-symmetry breaking—is a common ingredient across these families. This DMI unifies hallmark observations in parent compounds—incommensurate orders, spin-wave gaps, and noncollinear textures. Under hole doping, strong DMI drives spin vortices to merge with $\pi$-shifted hole stripes, forming hybrid vortex–hole stripe phases. These phases stabilize charge order while supporting, not suppressing, superconductivity. By contrast, under electron doping, these vortices pin holes and suppress long-range superconductivity. Our results establish DMI as a unifying link between noncollinear magnetism and superconductivity, identifying hole-stripe–vortex coupling as a microscopic pairing engine. Given that DMI is common across major superconductor families, these findings challenge the prevailing pairing mechanism and offer an experimentally testable roadmap for materials optimization.**

The mechanism of unconventional superconductivity—exhibited by cuprates (*1, 2*), iron pnictides and chalcogenides (*3-5*), and nickelates (*6-8*)—has remained elusive for decades. Unlike conventional Bardeen–Cooper–Schrieffer superconductors (*9*), these materials display high transition temperatures ($T_c$) that cannot be explained by electron–phonon coupling alone (*10-12*). A recurring empirical theme is that superconductivity often emerges when antiferromagnetic (AFM) order in parent compounds is suppressed by doping or applied external fields. This observation has led to a widespread hypothesis: spin fluctuations, generated by the emergence of nontrivial spin textures, may glue electrons into Cooper pairs (*13*), analogous to the role of phonons in conventional superconductors. Yet, despite extensive efforts, a unified understanding of how the diverse magnetic orders across different unconventional superconductors (UcSCs) contribute to the pairing mechanism remains conspicuously absent.

In the parent compounds of cuprates and iron pnictides/chalcogenides, collinear AFM order is generally believed to occur near $(\pi,\pi)$ (*14-19*). However, inelastic neutron scattering measurements on undoped detwinned samples and parent compounds reveal that spin fluctuations consistently deviate from $(\pi,\pi)$ by an incommensurability $\delta$, giving rise to incommensurate spin wave vectors with a detectable finite spin wave gap (*15, 20-24*). Strikingly, upon doping, this $\delta$ is found to be proportional to the $T_c$ in the extensively studied $La_{2-x}Sr_xCuO_4$(*25*) and $YBa_2Cu_3O_{6+x}$ (*26, 27*) families, up to optimal doping; meanwhile, a spin wave gap is universally observed (*15, 20-23*) across these UcSCs. These anomalies point to the possible role of noncollinear magnetic interactions. In nickelates—such as infinite-layer $NdNiO_2$, $La_3Ni_2O_7$, and $La_4Ni_3O_{10}$—the parent compounds exhibit $NiO_6$ octahedral rotations (*7, 28, 29*). The resulting structural asymmetry may profoundly influence the ground-state magnetism away from collinear AFM order. Direct evidence for noncollinear interactions comes from the discovery of topological spin vortexes in $YBa_2Cu_3O_{6.5}$, (*30, 31*) $Ba_{1-x}K_xFe_2As_2$ (*32, 33*) with different sizes as well as a hedgehog spin vortex crystal in $CaKFe_2As_2$ upon doping (*34*). These vortexes can generate integer-flux and fractional-flux which coexist with superconducitity. However, to our knowledge, no systematic investigation has either ruled out this possibility or elucidated how possible noncollinear magnetism or these spin textures might influence superconductivity, or indeed coexist with it below $T_c$.

Here, using first-principles calculations together with large-scale Monte-Carlo (MC) and density matrix renormalization group (DMRG) numerical simulations, we report a unvalued common feature, Dzyaloshinskii-Moriya interaction (DMI) (*35, 36*), in UcSCs. Remarkably, the above unexplained noncollinear magnetism related phenomena can be well understood by the revised Hamiltonian with the inclusion of DMI term. Under hole doping, strong DMI drives spin vortices to merge with π-phase-shifted hole stripes, self-organizing into hybrid vortex-hole structures that stabilize charge density waves (CDW) and protect the superconducting state. In contrast, under electron doping, the same spin vortices destroy the CDW and pin holes, effectively suppressing long-range superconducting order. Our results thus establish DMI as a universal bridge between noncollinear magnetism and superconductivity, and identify the mutual coupling between hole stripes and spin vortices as the essential microscopic origin of the superconducting pairing.

## Local inversion symmetry breaking induced DMI.

The atomistic structure of parent cuprates, exemplified by $YBa_2Cu_3O_6$ in Fig. 1A, consists of two $CuO_2$ layers per unit cell separated by yttrium and sandwiched between barium-oxygen layers. Although the overall crystal belongs to the centrosymmetric $D_{4h}$ point group, the local environment on the two sides of each $CuO_2$ plane is asymmetric—giving rise to the characteristic O–Cu–O buckling. This local inversion symmetry breaking is the key ingredient that naturally induces the Dzyaloshinskii–Moriya interaction (DMI) (*35, 36*). Following the Moriya rules(*36*), the DMI vector for nearest-neighbor Cu–Cu pair points perpendicular to the Cu–Cu bond (black arrows along $a$ in Fig. 1A). Moreover, the $C_{4z}$ rotational symmetry of the $CuO_5$ pyramid imposes a symmetric rotation pattern of the DMI vectors for the four equivalent nearest-neighbor bonds (lower panel of Fig. 1A).

In iron pnictides and chalcogenides, the crystal structure features two interpenetrating Fe$X$ ($X$ = P, As, Se, Te) sublattices within a single unit cell (see FeSe in Fig. 1B), with alkali, alkaline-earth, or rare-earth atoms and oxygen/fluorine serving as intercalation layers. Although the overall structure belongs to the centrosymmetric $D_{4h}$ point group—with an inversion center located between two Fe atoms from different sublattices—the $X$–Fe–$X$ buckling in each sublattice breaks local inversion symmetry. Consequently, though DMI is absent for first-nearest-neighbor Fe–Fe pairs due to global inversion, it emerges for second-nearest-neighbor pairs. Unlike the $C_{4z}$ rotational symmetry in cuprates, the $FeX_4$ tetrahedral geometry imposes $D_{2d}$ local symmetry, resulting in an asymmetric DMI (*37*) rotation pattern for the four second-nearest Fe–Fe bonds (lower panel of Fig. 1B). Nickelates exhibit a similar phenomenon. Despite belonging to different point groups—infinite-layer and Ruddlesden–Popper phases—all parent compounds display $NiO_6$ octahedral rotations. These rotations break local inversion symmetry and thereby induce DMI. Importantly, the specific rotation pattern of the $NiO_6$ octahedra dictates the orientation of the DMI vectors across different nickelate families. In the case of infinite-layer $NdNiO_2$, shown in Fig. 1C, the DMI is oriented out-of-plane. Collectively, these examples establish that local inversion symmetry breaking—whether through $CuO_5$ buckling, $FeX_4$ tetrahedral distortions, or $NiO_6$ octahedral rotations—universally generates DMI across diverse families of unconventional superconductors.

Thus, even in materials with global centro-symmetry, local structural distortions universally generate DMI—a feature that extends beyond cuprates to iron-based and nickelate superconductors. Yet, compared to well-studied phenomena such as CDW, spin density wave (SDW), pair density wave (PDW), pseudogap, strange metal, and Fermi liquid behavior (*10, 11, 38-43*), this noncollinear magnetic interaction has received little attention in superconductors (*44*). More generally, DMI favors perpendicular alignment between adjacent spins, inherently promoting noncollinear magnetic order. This natural consequence of DMI provides a crucial missing ingredient to address long-standing puzzles about the pairing mechanism that cannot be resolved within the conventional framework of collinear antiferromagnetism and spin fluctuations alone.

## DMI-induced incommensurate spin orders.

The structural analysis above leads to a central conclusion: the spin Hamiltonian of parent unconventional superconductors must include the DMI as an essential term:

$$H = \sum_{i \neq j} J_{ij} \cdot S_i \cdot S_j + \sum_{i \neq j} D_{ij} \cdot (S_i \times S_j) + A \sum_i (S_i \cdot \vec{\mathbf{z}})^2 \qquad (1)$$

where $S_i$ and $S_j$ are spins at sites $i$ and $j$, and $J_{ij}$, $D_{ij}$ and $A$ represent the isotropic Heisenberg exchange, the DMI vector, and the single-ion anisotropy, respectively. As summarized in Table I, all exchange interactions are AFM except for a tiny ferromagnetic $J_3$ in FeSe, indicating that AFM fluctuations dominate in the parent compounds. Crucially, DMI is present in every superconductor family examined. Notably, the ratio of the dominant DMI strength $D_{ij}$ to the leading exchange coupling $J_1$, *i.e.*, $D_{ij}/J_1$, lies between 0.05 and 0.08, a non-negligible value that guarantees a tangible noncollinear contribution. The calculated $A$ is comparatively very small. These calculated values are comparable to previous results (*19, 45-47*).

Mean-field theory, which restricted to commensurate magnetic orders, predicts a limited set of ground states for parent superconductors. However, it cannot account for incommensurate magnetic orders because it lacks noncollinear exchange interactions for classical spins (*42, 47*). Here we directly simulate the spin textures using Eq. (1) via Monte Carlo (MC) simulations under a 50×50 supercell. For $YBa_2Cu_3O_6$ (Fig. 2A), one $CuO_2$ layer exhibits a spin cycloid AFM structure. Its Fourier transform (Fig. 2B) reveals a spin fluctuation peak at $\mathbf{Q} \sim (0.5 \pm \delta, 0.5)$ in the square lattice (one Cu per unit cell), with $\delta = 0.02$, a clear deviation from the Néel order at (0.5, 0.5), demonstrating incommensurate magnetism. Similarly, monolayer FeSe shows a single-stripe helix spin texture (Fig. 2C), with the corresponding reciprocal peak at $\mathbf{Q} \sim (0.5 \pm \delta, 0.5 \pm \delta)$ in the tetragonal lattice (two Fe per unit cell), also with $\delta = 0.02$ (Fig. 2D). In striking contrast, $NdNiO_2$ (Fig. 2E) exhibits hardly any spin tilting or rotation, because both the single-ion anisotropy $A$ and the DMI vector are oriented out-of-plane. This results in a conventional Néel spin fluctuation peak at $\mathbf{Q} = (0.5, 0.5)$ (Fig. 2F). Notably, the incommensurate peaks $(0.5 \pm \delta, 0.5 \pm \delta)$ observed in tetragonal FeSe can be decomposed into two Néel magnetic vectors $(0.5 \pm \delta, 0.5 \pm \delta)$ for the two interpenetrating square Fe sublattices (see fig. S1), revealing a common sublattice-based magnetic structure across cuprates, nickelates, and iron-based superconductors.

The DMI naturally promotes site-dependent spin rotations, thereby generating spin incommensurability. Indeed, when the DMI is artificially removed from the Hamiltonian, the incommensurability in both $YBa_2Cu_3O_6$ and FeSe disappears entirely (see fig. S2). We note, however, that experiments on cuprates and iron pnictides/chalcogenides typically report four spin fluctuation peaks, whereas our simulations yield only two. We attribute this discrepancy to twinning in the measured samples (*24, 48*), a common feature in unconventional superconductors and the randomness of MC simulation. Separately, in the underdoped regime, theoretical and numerical studies have suggested that doped holes can self-organize into stripes within an AFM background, potentially contributing to incommensurate magnetic orders. Nevertheless, we emphasize that the incommensurate order observed in the parent compounds requires an intrinsic mechanism. Our results demonstrate that noncollinear magnetic exchange interactions—specifically DMI—are both sufficient and necessary to explain these incommensurate orders, which are therefore an intrinsic property of parent compounds of $YBa_2Cu_3O_6$ and FeSe.

Figs. 2G-2I present the spin wave dispersions for the three compounds, calculated using linear spin wave theory. In DMI-free $YBa_2Cu_3O_6$ (blue curve, Fig. 2G), the lowest spin excitations arise from Néel order at (1/2, 1/2) and ferromagnetic order at (1, 0). Introducing DMI opens a spin wave gap $E_g = 26.98$ meV (red curve, Fig. 2G), which falls in the experimental values (*27*). Moreover, the lowest excitation now appears at a wave vector close to, but not exactly at, (1/2, 1/2), consistent with both experimental observations and our MC results (Fig. 2B). Intensity cuts through the spin wave spectrum in the energy range 20–30 meV reveal that the isotropic spin excitation becomes $C_2$-symmetric upon inclusion of DMI (fig. S3), indicating the emergence of spin nematicity (*49*). A spin wave gap $E_g = 21.52$ meV and spin nematicity are also observed in FeSe when the asymmetric DMI is included (Fig. 2H and fig. S4) which matches well to experiments (*50, 51*). Notably, the spin gap at the single-stripe order (1/2, 1/2) is only 3.5 meV lower than that at the Néel order (1, 0), suggesting competing magnetic orders—a feature confirmed by inelastic neutron scattering measurements on FeSe (*50*).

For the infinite-layer $NdNiO_2$, out-of-plane AFM coupling $J_{\text{oop}}$ is also considered, resulting in four branches in the spin wave spectrum (Fig. 2I). This interlayer coupling opens a gap of ~9.7 meV between layers. Additionally, DMI induces a tiny spin wave gap $E_g = 1.13$ meV at the Néel order (1, 0) (Fig. 2I and fig. S5). Such a small gap can hardly stabilize the ground state magnetic order against thermal fluctuations and crystal defects, which may explain why spin stripe orders exhibit the strongest spin fluctuations in $NdNiO_2$ (*52*) and $La_3Ni_2O_7$ (*53, 54*), as observed by resonant inelastic X-ray scattering and resonant soft X-ray scattering. Owing to the weak magnetic exchange interactions and small magnetic moment in $NdNiO_2$, the spin wave dispersion extends only to about 100 meV—roughly half of that in $YBa_2Cu_3O_6$ and FeSe. Finally, the isotropic out-of-plane DMI in $NdNiO_2$ does not induce $C_2$-symmetric spin excitations (fig. S5), in contrast to the cuprates and iron-based cases.

**Noncollinear spin texture phase diagram.**

The classical mean-field phase diagram based on Heisenberg-type exchange interactions alone has been known for over a decade, yet it cannot explain the noncollinear spin vortices observed in the underdoped regime of unconventional superconductors. To gain deeper insight, we perform MC simulations on a 100×100 supercell system at low temperatures (~1 K) and map out the typical spin texture phase diagram as a function of $J_1, J_2, J_{\text{inter}}$, and $D$. For cuprates, we fix the AFM $J_1 = 50$ meV and vary the other parameters, considering both FM and AFM interlayer couplings $J_{\text{inter}}$. When $J_{\text{inter}}$ is not too strong, the phase diagram (Fig. 3A) shows little change, indicating quasi-2D behavior. For clarity, we present the case $J_{\text{inter}}/J_1 = 0$ in Fig. 3B.

Besides the Néel state at $\mathrm{Q} = (1/2,1/2)$ (Fig. 3C) in the lower-left region and the single-stripe magnetic order at $\mathrm{Q} = (1/2,0)$ for $J_2/J_1 \gtrsim 0.6$ (Fig. 3D), three intermediate phases appear between the Néel and single-stripe states. Compared with the Néel state, the single-stripe state can be viewed as a close-packed array of $\pi$-phase-shift domain walls aligned along the *y*-axis. The intermediate phase is pure spin spiral (Fig. 3E), where the $\pi$-phase-shift domain walls are not close-packed, giving an incommensurate SDW with $\mathrm{Q} = (1/2, q_y)$, $0 < q_y < 1/2$. Most importantly, topological spin vortex states (Figs. 3F and 3G) emerge in two distinct regimes: a small-vortex state with a stripe domain near the spiral region, and a large-vortex state with a broad AFM domain near the Néel region. In the small-vortex state, vortexes reside within a stripe domain, cutting it into segments—a structure we term a vortex-decorated domain, or in quantum language, "bubbles" in the domain. These two vortex states closely resemble the topological spin textures experimentally observed in $YBa_2Cu_3O_{6+x}$ (*30*). Moreover, the mean-field theory suggests $J_3/J_1$=0.04 which prohibits the vortex formation from competing magnetic frustration and leave the DMI to be the primary origin. In iron-based superconductors, though the type of DMI is different from that of cuprates, vortexes can also emerge (fig. S6).

Thus, our calculated $J_1$–$J_2$–$J_{\text{inter}}$–$D$ magnetic phase diagram not only reproduces incommensurate magnetic orders but also reveals topological spin vortexes. This demonstrates that noncollinear magnetic interactions are essential for capturing the complex spin textures in these superconductors, particularly the interplay between spin vortexes and stripe domains. Upon hole doping, holes preferentially occupy lattice sites and form stripes in these stripe domains, lowering the ground-state energy. In the absence of DMI, longer stripes are energetically favored. However, the introduction of DMI generates vortexes that compete with stripe formation: vortexes seek to maintain noncollinear spin textures, whereas holes tend to stabilize stripes to support their own motion. The resulting hole stripes exhibit a finite stiffness against spin textures between domain walls, giving rise to an intertwined relationship between SDW and CDW orders. This competition can be visualized by tuning the relative sizes of spin vortexes and hole stripes, which is intimately controlled by $J_1$–$J_2$–$D$.

## Effects of DMI on stripe phase.

The superconductivity of cuprates can be described by the single-band $t$-$t'$-$U$ Hubbard model (*55-59*). Its Hamiltonian reads:

$$\mathcal{H}_H = -t \sum_{<i,j>,\sigma} \hat{c}^{\dagger}_{i,\sigma}\, \hat{c}_{j,\sigma} - t' \sum_{\ll i,j \gg,\sigma} \hat{c}^{\dagger}_{i,\sigma}\, \hat{c}_{j,\sigma} + U \sum_i \hat{n}_{i,\uparrow}\, \hat{n}_{j,\downarrow} \tag{2}$$

where $\hat{c}^{\dagger}_{i,\sigma}$ ($\hat{c}_{i,\sigma}$) creates (annihilates) an electron at site $i$ with spin $\sigma = \{\uparrow,\downarrow\}$, and $\hat{n}_{i,\sigma} = \hat{c}^{\dagger}_{i,\sigma}\hat{c}_{i,\sigma}$ is the number operator. The sums $\langle i,j \rangle$ and $\ll i,j \gg$ run over nearest-neighbor and next-nearest-neighbor sites, respectively. In cuprates, the next-nearest hopping coefficient is typically negative ($t' < 0$), which corresponds to hole doping in the following analysis. The electron-doped case can be studied by a particle–hole transformation, which effectively reverses the sign of $t'$. In this work, we fix the ratio of the nearest-neighbor hopping coefficient $t$ to the on-site repulsion strength $U$ at $U/t = 12$.

Because the DMI originates from SOC in systems with broken inversion symmetry, we include a Rashba-type SOC term in the cuprates to explore its influence on superconductivity, simulating the influence of DMI:

$$\mathcal{H}_{\text{SOC}} = i\lambda \sum_{i,\sigma,\sigma'} \left[\hat{c}^{\dagger}_{i,\sigma}\hat{\sigma}^{y}_{\sigma,\sigma'}\hat{c}_{i+\hat{x},\sigma'} + \hat{c}^{\dagger}_{i,\sigma}\hat{\sigma}^{x}_{\sigma,\sigma'}\hat{c}_{i+\hat{y},\sigma'} + \text{h.c.}\right] \tag{3}$$

Here, $\hat{\sigma}^{\mu}_{\sigma,\sigma'}$ ($\mu = x,y,z$) are the Pauli matrix elements, and $i + \hat{x}$ ($i + \hat{y}$) denotes the nearest site along the $\hat{x}$ ($\hat{y}$) direction on the square lattice. The total Rashba–Hubbard Hamiltonian is then:

$$\mathcal{H} = \mathcal{H}_H + \mathcal{H}_{\text{SOC}} \tag{4}$$

To investigate the superconducting properties of the Rashba–Hubbard model, we perform large-scale DMRG simulations (*60-62*). We employ cylindrical geometries with typical dimensions $L_x = 12$ and $L_y = 6$ (or comparable sizes). The doping level in Fig. 4 is fixed at hole concentration $\delta_h = 1/9$ in the optimally doping regime. By modulating the periodic spatial distribution of doped holes, CDW stripe phases have been widely observed in studies of the Hubbard model (*61-63*). In general, different stripe phases can be characterized by distinct filling fractions $f = \delta_h \lambda_c$, which correspond to CDW order with various wavelengths $\lambda_c$ (taking the lattice spacing as unity). To better visualize the spin textures in the DMRG results, the AFM background is subtracted (see **Methods**).

Fig. 4A displays five main phases classified by CDW and SDW orders. We label the intermediate region without clearly definable features as a phase-separation (PS) regime, while the other four phases exhibit distinguishable characteristics. Interestingly, when $\lambda = 0$ and $t' \lesssim -0.4$, the PS region gives way to a possible Wigner crystal (WC) phase. In the hole-doped regime ($t' < 0$), weak SOC yields a CDW with filling fraction $f = 2/3$, consistent with the SOC-free case (*57*). However, in contrast to the SOC-free limit, including SOC forces the original AFM order to evolve into a spiral SDW—a staggered spiral magnetic pattern in the $xz$ plane (Fig. 4B). Remarkably, the magnetic moments on the two sides of each stripe retain a $\pi$-phase shift in both the $x$ and $z$ spin components. This result is consistent with the anticipated outcome obtained upon doping the large-vortex state in the parent cuprates (Fig. 3F), wherein hole stripes emerge within a broad AFM domain and spin textures are corresponding rearranged between stripes. Consequently, within a narrow window adjacent to each hole stripe, an $xz$-plane spiral SDW becomes discernible.

As SOC increases, a vortex-like spin configuration with quantized flux $1/2$ emerges (Fig. 4C). Importantly, the vortex centers are spinless—fully consistent with the stripe structure, which itself carries no net spin polarization. Thus, these fractional spin vortexes do not disrupt the stripes; instead, the two coexist naturally, and the $\pi$-phase shift across stripes remains robust. Because doped holes are pinned at the vortex centers, the vortex size must be commensurate with the CDW wavelength. In Fig. 4C, four $3 \times 3$ vortexes appear within the selected region, indicating a change in CDW periodicity from the CDW-$2/3$ phase to a CDW-$1/3$ phase, effectively doubling the number of stripes. We denote this phase as hole-CDW-1/3. This behavior can be well understood by doping the small-vortex state in the parent cuprates (Fig. 3H left), wherein a hole stripe forms along a stripe domain and the holes localize within the cores of fractional spin vortexes.

Upon further increasing SOC (typically $\lambda \gtrsim 0.4$), the strong DMI favors perpendicular alignment of neighboring spins, destroying the spin vortexes and giving rise to a spiral magnetic order in the $xz$ plane (Fig. 4D). This spiral texture preserves a $\pi$-phase shift for the $x$ component. After restoring the staggered AFM background (i.e., removing a site-dependent phase factor), Fig. 4D reveals that spins exhibit AFM alignment along the lattice $y$-direction—including on the stripes, where net spin polarization remains finite. Along the $x$-direction, the spins display a three-period modulation, reminiscent of behavior found in 1D spin models with strong DMI (*64*). This three-period spiral matches the CDW wavelength, allowing SDW and CDW orders to coexist, and the CDW filling fraction remains $1/3$.

The electron-doped regime exhibits qualitatively different behavior. Because the $\pi$-phase shift is decoupled from the hole stripe, weak SOC still induces a spiral SDW order, but the SDW wavelength becomes incommensurate with that of the CDW (Fig. 4E)—a clear signature of order separation. Different from the hole-doped regime that hosts a CDW-$2/3$ phase, we naturally obtain a CDW with filling fraction $f = 1/3$; we label this elec-CDW-1/3 phase in the phase diagram. Upon further increasing SOC, vortex-like magnetic textures also emerge, and holes become pinned at the vortex cores. More interestingly, spin vortexes unexpectedly disrupt hole stripes in both PS and Vortex (Fig. 4F) regimes, compared to the hole-CDW-$1/3$ phase in the corresponding hole-doped regime (Fig. 4D).

Overall, weak SOC leads to a CDW-$2/3$ phase in the hole-doped regime and an elec-CDW-$1/3$ phase in the electron-doped regime, akin to the SOC-free cases. When $L_y/2$ is an even integer, the fill fractions in both regimes become $f = 1/2$ (e.g., $L_y = 4$ in fig. S7). Neverthess, when fractional spin vortexes are induced by relatively strong SOC, the two regimes diverge markedly. In the hole-doped regime, the fractional spin vortices remain strongly coulpled to the $\pi$-phase-shifted hole stripes, forming hybrid vortex-hole stripes and generating the new hole-CDW-$1/3$ phase. In stark contrast, in the electron-doped regime, the absence of a $\pi$-phase shift that topologically protects hole stripes renders these emerged spin vortexes destabilizing to the CDW in both PS and Vortex regimes. These diverse characteristics give rise to distinct effects on the superconducting porperties.

## Effects of DMI on superconductivity.

In Fig. 5, we further compute the ground state of a longer cylinder using a suggested parameter set for $YBa_2Cu_3O_6$. The hole distribution and the spin textures (Fig. 5A) confirm the existence of a stable CDW with a filling fraction of $1/3$, where each hole stripe contains a pair of fractional spin vortexes, matching the characteristics of the hole-CDW- $1/3$ phase. Furthermore, to investigate the superconducting properties, we compute the spin-singlet superconducting correlation (SSC) functions for the ground state (Fig. 5B, see details in **Methods**). The SSC exhbits a clear power-law decay at long distances along the lattice $x$-direction indicating quasi-long-range superconducting order, a behavior consistent with expectations for 2D superconductivity on a quasi-1D cylindrical lattice. In DMRG studies, such power-law decay is a necessary criterion for identifying superconducting order. In the CDW-$2/3$ phase on the hole-doped side (fig. S8), the SSC also exhibits a clear power-law decay at long distances, confirming the existence of superconductivity, although the CDW wavelength doubles as the filling fraction changes from $2/3$ to $1/3$. In stark contrast, on the electron-doped side, the SOC-induced spin vortexes destroy CDW and the corresponding SSC are suppressed at long distances (fig. S8) in the Vortex phase.

From the above results, we identify two distinct superconducting phases in the hole-doped regime. To clarify the superconducting pairing structures in these two phases, we compute the pair-pair correlation matrix $C_{m,n}(d)$ between two vertical bonds separated by a distance $d$ (Fig. 5A). Its matrix elements are defined as:

$$C_{m,n} \equiv C_{(\sigma_1,\sigma_2);(\sigma_3,\sigma_4)} = \langle \hat{c}^\dagger_{\sigma_1,i} \hat{c}^\dagger_{\sigma_2,i+\hat{y}} \hat{c}_{\sigma_3,j} \hat{c}_{\sigma_4,j+\hat{y}} \rangle \tag{5}$$

where $m, n \in [1,4]$ correspond to the four spin configurations $\{\uparrow\uparrow, \uparrow\downarrow, \downarrow\uparrow, \downarrow\downarrow\}$ for both $(\sigma_1, \sigma_2)$ and $(\sigma_3, \sigma_4)$. To extract the dominant pairing symmetry, we perform a singular value decomposition (SVD):

$$C_{(\sigma_1,\sigma_2);(\sigma_3,\sigma_4)} = U_{(\sigma_1,\sigma_2);\alpha} \, \Lambda_\alpha \, (V^\dagger)_{\alpha;(\sigma_3,\sigma_4)} \tag{6}$$

and identify a singular vector pair associated with the largest singular value as the dominant pairing modes on the two vertical bonds. For each of these singular vectors, we evaluate the expectation value of the square of the total-spin operator $\langle S^2 \rangle$, yielding $\langle S^2 \rangle_{\sigma_1,\sigma_2}$ and $\langle S^2 \rangle_{\sigma_3,\sigma_4}$, shown in Fig. 5C. The results reveal that, compared with the SOC-free case, the superconducting order in the hole-CDW-$1/3$ phase is no longer dominated by a pure singlet channel. Moreover, as the bond distance $d$ increases, $\langle S^2 \rangle_{\sigma_1,\sigma_2}$ converges to a stable value, indicating that the dominant pairing is neither pure singlet nor pure triplet but rather a mixed singlet–triplet configuration—a qualitative change induced by SOC. In addition, $\langle S^2 \rangle_{\sigma_3,\sigma_4}$ exhibits the same periodic oscillation as the underlying CDW, demonstrating that the angular momentum distribution of superconducting pairing is spatially inhomogeneous. Specifically, the pairing angular momentum is smaller within the hole stripes, whereas it becomes relatively larger at sites on their right side. However, the small yet nonzero angular momentum of the paired holes presents a significant challenge for detection in experiments.

Overall, it is evident that a stable CDW phase is essential for the emergence of superconductivity in the Rashba–Hubbard model. In the hole-doped case, the presence of either $\pi$-phase-shifted hole stripes under weak SOC or hybrid vortex-hole stripes under strong SOC allows for the coexistence of CDW and superconductivity. While in the electron-doped regime, the emerged spin vortices disrupt the CDW order and suppresses the superconductivity. The intertwining of these magnetic patterns with hole stripes engenders distinct superconducting responses across the two regimes. Moreover, SOC significantly alters the pairing symmetry in the hole-doped regime, giving rise to a mixed singlet–triplet pairing character, rather than the purely singlet pairing realized in the doped colinear AFM background.

**Discussion**

From crystal symmetry analysis, we find that the antisymmetric DMI is a common feature across the unconventional superconductors studied here—including cuprates, iron pnictides/chalcogenides, and nickelates in their parent compounds—as well as in other superconductors such as heavy-fermion metals, Kagome crystals (*65*), and CrAs (*66*). DMI thus generically promotes noncollinear magnetism in high-temperature superconductors. Interestingly, spin-canted order (*67*) may also contribute to the pairing interaction in Bernal-stacked bilayer graphene (*68*) and rhombohedral trilayer graphene (*69*). However, to date, noncollinear magnetism has received significant attention only in $La_2CuO_4$ (*70*). The key questions for future work are whether this antisymmetric DMI can be experimentally detected in other compounds and how it further influences the pairing mechanism.

In the underdoped regime, scanning tunneling microscopy measurements have already detected segmented stripes (*71, 72*). Our study provides an alternative explanation for why these stripes are cut into segments in the absence of impurity effects: spin vortexes puncturing the stripe in the small-vortex state. Notably, at parameter values $t'/t \approx -0.45$ and $\lambda/t \approx 0.15$, close to those suggested by first-principles calculations (see details in **Methods**) and the fitting of the spin-wave gap, the CDW-$2/3$, hole-CDW-$1/3$, and potential WC phases compete strongly. Their pairing angular momentum also remains close to that of a pure singlet pairing scenario, making it difficult to determine which state corresponds to experimental observations. Nevertheless, based on our simulations, we propose that future experiments should simultaneously measure the lengths of hole stripes, as the ratio between spin vortex sizes and hole stripe lengths may provide valuable insights.

In summary, we identify a previously unvalued common feature across unconventional superconductors, the antisymmetric DMI. By including this noncollinear magnetic interaction, the diverse noncollinear magnetism observed in parent compounds, including incommensurate magnetic order, spin wave gaps, and spin vortexes, can be well described. Furthermore, we elucidate the influence of DMI on superconductivity. In the hole-doped regime, SDW and CDW orders remain intertwined and preserve long-range superconductivity. In contrast, in the electron-doped regime, holes become pinned at the centers of DMI-induced vortices, which destroys CDW and suppresses superconductivity. These results collectively indicate that the pairing mechanism in unconventional superconductors requires fundamental revision.

**ACKNOWLEDGEMENTS**

**Funding:** This work is supported by Science Challenge Project (Grant No. TZ2025013), MOST (Grants No. 2022YFA1402700) and NSFC (Grant Nos. W2511008, 12174020). Technical and human supports provided by Tianhe2-JK at CSRC and DIPC Supercomputing Center are gratefully acknowledged. **Author contributions:** B.Y., S. H., H.Q.L. and B.H. conceived and led the project. B. Y. did the DFT and MC simulations. Y. C. and S. H. did the DMRG calculations. B.Y., Y. C., S. H., and B.H. prepared the manuscript. All authors discussed the results and contributed to the manuscript. **Competing interests:** The authors declare no competing interests.

## REFERENCES AND NOTES

1. J. G. Bednorz, K. A. Müller, Possible high Tc superconductivity in the Ba−La−Cu−O system. *Z. Phys. B* **64**, 189-193 (1986).
2. C. Tsuei, J. Kirtley, Pairing symmetry in cuprate superconductors. *Rev. Mod. Phys.* **72**, 969 (2000).
3. Y. Kamihara, T. Watanabe, M. Hirano, H. Hosono, Iron-based layered superconductor $La[O_{1-x}F_x]FeAs$(x= 0.05− 0.12) with *Tc* = 26 K. *J. Am. Chem. Soc.* **130**, 3296-3297 (2008).
4. Mazin, II, Superconductivity gets an iron boost. *Nature* **464**, 183-186 (2010).
5. G. R. Stewart, Superconductivity in iron compounds. *Rev. Mod. Phys.* **83**, 1589-1652 (2011).
6. D. Li *et al.*, Superconductivity in an infinite-layer nickelate. *Nature* **572**, 624-627 (2019).
7. H. Sun *et al.*, Signatures of superconductivity near 80 K in a nickelate under high pressure. *Nature* **621**, 493-498 (2023).
8. Q. Gu, H. H. Wen, Superconductivity in nickel-based 112 systems. *Innovation* **3**, 100202 (2022).
9. J. Bardeen, L. N. Cooper, J. R. Schrieffer, Theory of Superconductivity. *Phys. Rev.* **108**, 1175-1204 (1957).
10. D. J. Scalapino, A common thread: The pairing interaction for unconventional superconductors. *Rev. Mod. Phys.* **84**, 1383-1417 (2012).
11. M. R. Norman, The Challenge of Unconventional Superconductivity. *Science* **332**, 196-200 (2011).
12. G. R. Stewart, Unconventional superconductivity. *Adv. Phys.* **66**, 75-196 (2017).
13. P. Dai, J. Hu, E. Dagotto, Magnetism and its microscopic origin in iron-based high-temperature superconductors. *Nat. Phys.* **8**, 709-718 (2012).
14. R. Coldea *et al.*, Spin waves and electronic interactions in $La_2CuO_4$. *Phys. Rev. Lett.* **86**, 5377-5380 (2001).
15. H. A. Mook *et al.*, Magnetic order in $YBa_2Cu_3O_{6+x}$ superconductors. *Phys. Rev. B* **66**, 144513 (2002).
16. N. S. Headings, S. M. Hayden, R. Coldea, T. G. Perring, Anomalous high-energy spin excitations in the high-Tc superconductor-parent antiferromagnet $La_2CuO_4$. *Phys. Rev. Lett.* **105**, 247001 (2010).
17. M. K. Chan *et al.*, Commensurate antiferromagnetic excitations as a signature of the pseudogap in the tetragonal high-Tc cuprate $HgBa_2CuO_{4+\delta}$. *Nat. Commun.* **7**, 10819 (2016).
18. C. de la Cruz *et al.*, Magnetic order close to superconductivity in the iron-based layered $LaO_{1-x}F_xFeAs$ systems. *Nature* **453**, 899-902 (2008).
19. F. Ma, W. Ji, J. Hu, Z. Y. Lu, T. Xiang, First-principles calculations of the electronic structure of tetragonal α-FeTe and α-FeSe crystals: evidence for a bicollinear antiferromagnetic order. *Phys. Rev. Lett.* **102**, 177003 (2009).
20. W. Bao *et al.*, Tunable (δπ, δπ)-type antiferromagnetic order in α-Fe(Te,Se) superconductors. *Phys. Rev. Lett.* **102**, 247001 (2009)..
21. B. Wells *et al.*, Incommensurate spin fluctuations in high-transition temperature superconductors. *Science* **277**, 1067-1071 (1997).
22. D. K. Pratt *et al.*, Incommensurate spin-density wave order in electron-doped $BaFe_2As_2$ superconductors. *Phys. Rev. Lett.* **106**, 257001 (2011).
23. S. Wan *et al.*, Direct visualization of a static incommensurate antiferromagnetic order in Fe-doped $Bi_2Sr_2CaCu_2O_{8+\delta}$. *Proc. Natl. Acad. Sci. U.S.A.* **118**, e2115317118 (2021).
24. T. Chen *et al.*, Anisotropic spin fluctuations in detwinned FeSe. *Nat. Mater.* **18**, 709-716 (2019).
25. K. Yamada *et al.*, Doping dependence of the spatially modulated dynamical spin correlations and the superconducting-transition temperature in $La_{2-x}Sr_xCuO_4$. *Phys. Rev. B* **57**, 6165 (1998).
26. A. Balatsky, P. Bourges, Linear Dependence of Peak Width in χ(q,ω) vs Tc for $YBa_2Cu_3O_{6+x}$ Superconductors. *Phys. Rev. Lett.* **82**, 5337 (1999).
27. P. Dai, H. A. Mook, R. D. Hunt, F. Doğan, Evolution of the resonance and incommensurate spin fluctuations in superconducting $YBa_2Cu_3O_{6+x}$. *Phys. Rev. B* **63**, 054525 (2001).
28. C. Zhang, X. Cai, C.-X. Zhang, S.-H. Wei, H.-X. Deng, Antiferromagnetism driven charge density wave in infinite-layer $NdNiO_2$ nickelates. *Phys. Rev. B* **109** (2024).
29. Y. Zhu *et al.*, Superconductivity in pressurized trilayer $La_4Ni_3O_{10-\delta}$ single crystals. *Nature* **631**, 531-536 (2024).
30. Z. Wang *et al.*, Topological spin texture in the pseudogap phase of a high-Tc superconductor. *Nature* **615**, 405-410 (2023).
31. J. R. Kirtley *et al.*, Direct Imaging of Integer and Half-Integer Josephson Vortices in High- Tc Grain Boundaries. *Phys. Rev. Lett.* **76**, 1336-1339 (1996).
32. Y. Iguchi *et al.*, Superconducting vortices carrying a temperature-dependent fraction of the flux quantum. *Science* **380**, 1244-1247 (2023).
33. Y. Zheng *et al.*, Observation of quantum vortex core fractionalization and skyrmion formation in a superconductor. *Science* **393**, 80-84 (2026).
34. W. R. Meier *et al.*, Hedgehog spin-vortex crystal stabilized in a hole-doped iron-based superconductor. *npj Quant. Mater.* **3**, 5 (2018).
35. I. DZYALOSHINSKY, A thermodynamic theory of "weak" ferromagnetism of antiferromagnetics. *J. Phys. Chem. Solids* **4**, 241-255 (1958).
36. T. Moriya, Anisotropic Superexchange Interaction and Weak Ferromagnetism. *Phys. Rev.* **120**, 91-98 (1960).
37. A. K. Nayak *et al.*, Magnetic antiskyrmions above room temperature in tetragonal Heusler materials. *Nature* **548**, 561-566 (2017).
38. Q. Gu *et al.*, Detection of a pair density wave state in $UTe_2$. *Nature* **618**, 921-927 (2023).

39. A. Aishwarya *et al.*, Melting of the charge density wave by generation of pairs of topological defects in UTe2. *Nat. Phys.* **20**, 964-969 (2024).
40. D. F. Agterberg *et al.*, The physics of pair-density waves: cuprate superconductors and beyond. *Annu. Rev. Condens. Matter Phys* **11**, 231-270 (2020).
41. B. Keimer, S. A. Kivelson, M. R. Norman, S. Uchida, J. Zaanen, From quantum matter to high-temperature superconductivity in copper oxides. *Nature* **518**, 179-186 (2015).
42. D. C. Johnston, The puzzle of high temperature superconductivity in layered iron pnictides and chalcogenides. *Adv. Phys.* **59**, 803-1061 (2010).
43. R. M. Fernandes *et al.*, Iron pnictides and chalcogenides: a new paradigm for superconductivity. *Nature* **601**, 35-44 (2022).
44. T. Thio *et al.*, Antisymmetric exchange and its influence on the magnetic structure and conductivity of $La_2CuO_4$. *Phys. Rev. B* **38**, 905-908 (1988).
45. X. Wan, T. A. Maier, S. Y. Savrasov, Calculated magnetic exchange interactions in high-temperature superconductors. *Phys. Rev. B* **79**, 155114 (2009).
46. H.-Y. Cao, S. Chen, H. Xiang, X.-G. Gong, Antiferromagnetic ground state with pair-checkerboard order in FeSe. *Phys. Rev. B* **91**, (2015).
47. J. K. Glasbrenner *et al.*, Effect of magnetic frustration on nematicity and superconductivity in iron chalcogenides. *Nat. Phys.* **11**, 953-958 (2015).
48. Q. Wang *et al.*, Strong interplay between stripe spin fluctuations, nematicity and superconductivity in FeSe. *Nat. Mater.* **15**, 159-163 (2016).
49. D. Vaknin, Magnetic nematicity: A debated origin. *Nat. Mater.* **15**, 131-132 (2016).
50. Q. Wang *et al.*, Magnetic ground state of FeSe. *Nat. Commun.* **7**, 12182 (2016).
51. X. Lu *et al.*, Spin-excitation anisotropy in the nematic state of detwinned FeSe. *Nat. Phys.* **18**, 806-812 (2022).
52. H. Lu *et al.*, Magnetic excitations in infinite-layer nickelates. *Science* **373**, 213-216 (2021).
53. X. Chen *et al.*, Electronic and magnetic excitations in La(3)Ni(2)O(7). *Nat Commun* **15**, 9597 (2024).
54. N. K. Gupta *et al.*, Anisotropic spin stripe domains in bilayer $La_3Ni_2O_7$. *Nat. Commun.* **16**, 6560 (2025).
55. C.-M. Chung *et al.*, Plaquette versus ordinary *d*-wave pairing in the t′-Hubbard model on a width-4 cylinder. *Phys. Rev. B* **102**, 041106 (2020).
56. Y.-F. Jiang, J. Zaanen, T. P. Devereaux, H.-C. Jiang, Ground state phase diagram of the doped Hubbard model on the four-leg cylinder. *Phys. Rev. Research* **2**, 033073 (2020).
57. Y.-F. Jiang, T. P. Devereaux, H.-C. Jiang, Ground-state phase diagram and superconductivity of the doped Hubbard model on six-leg square cylinders. *Phys. Rev. B* **109**, 085121 (2024).
58. S. R. White, Density matrix formulation for quantum renormalization groups. *Phys. Rev. Lett.* **69**, 2863 (1992).
59. S. R. White, Density-matrix algorithms for quantum renormalization groups. *Phys. Rev. B* **48**, 10345 (1993).
60. U. Schollwöck, The density-matrix renormalization group. *Rev. Mod. Phys.* **77**, 259-315 (2005).
61. E. W. Huang, C. B. Mendl, H.-C. Jiang, B. Moritz, T. P. Devereaux, Stripe order from the perspective of the Hubbard model. *npj Quant. Mater.* **3**, 22 (2018).
62. B.-X. Zheng *et al.*, Stripe order in the underdoped region of the two-dimensional Hubbard model. *Science* **358**, 1155-1160 (2017).
63. H. Xu *et al.*, Coexistence of superconductivity with partially filled stripes in the Hubbard model. *Science* **384**, eadh7691 (2024).
64. Y. Chu, S. Hu, T. Wang, Realization of a period-3 coplanar state in one-dimensional spin-orbit-coupled optical lattices. *Phys. Rev. A* **111**, L011304 (2025).
65. S. D. Wilson, B. R. Ortiz, $AV_3Sb_5$ kagome superconductors. *Nat. Rev. Mater.* **9**, 420-432 (2024).
66. W. Wu *et al.*, Superconductivity in the vicinity of antiferromagnetic order in CrAs. *Nat. Commun.* **5**, 5508 (2014).
67. Z. Dong, É. Lantagne-Hurtubise, J. Alicea, Superconductivity from spin-canting fluctuations in rhombohedral graphene. *Phys. Rev. X* **16**, 011055 (2026).
68. Y. Zhang *et al.*, Enhanced superconductivity in spin–orbit proximitized bilayer graphene. *Nature* **613**, 268-273 (2023).
69. C. L. Patterson *et al.*, Superconductivity and spin canting in spin–orbit-coupled trilayer graphene. *Nature*, **641**, 632–638 (2025).
70. T. Thio, A. Aharony, Weak ferromagnetism and tricriticality in pure $La_2CuO_4$. *Phys. Rev. Lett.* **73**, 894-897 (1994).
71. H. Li *et al.*, Low-energy gap emerging from confined nematic states in extremely underdoped cuprate superconductors. *npj Quant. Mater.* **8**, 18 (2023).
72. S. Ye *et al.*, Visualizing the Zhang-Rice singlet, molecular orbitals and pair formation in cuprate. Preprint at https://arxiv.org/abs/2309.09260.

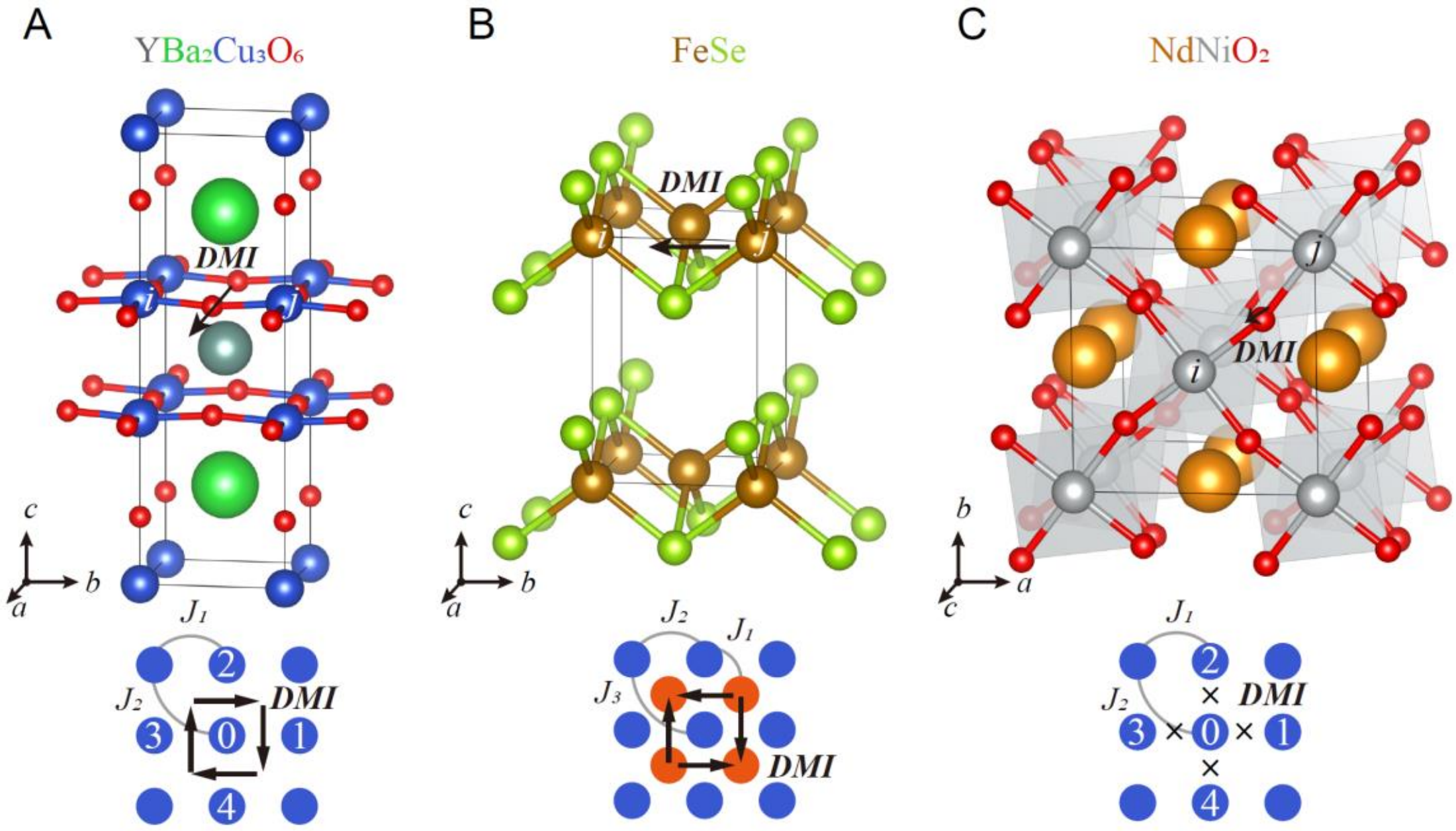

**Fig. 1. Local inversion symmetry breaking induces DMI across unconventional superconductors.** Crystal structures of **a**, $YBa_2Cu_3O_6$, **b**, FeSe, and **c**, $NdNiO_2$. Black arrows show DMI vectors for representative magnetic atom pairs. Lower panels depict the DMI vector patterns on the four relevant bonds. Blue and red circles indicate different sublattices. Local buckling, tetrahedral distortions, or octahedral rotations break local inversion symmetry, generating DMI in cuprates, iron-based superconductors, and nickelates, respectively.

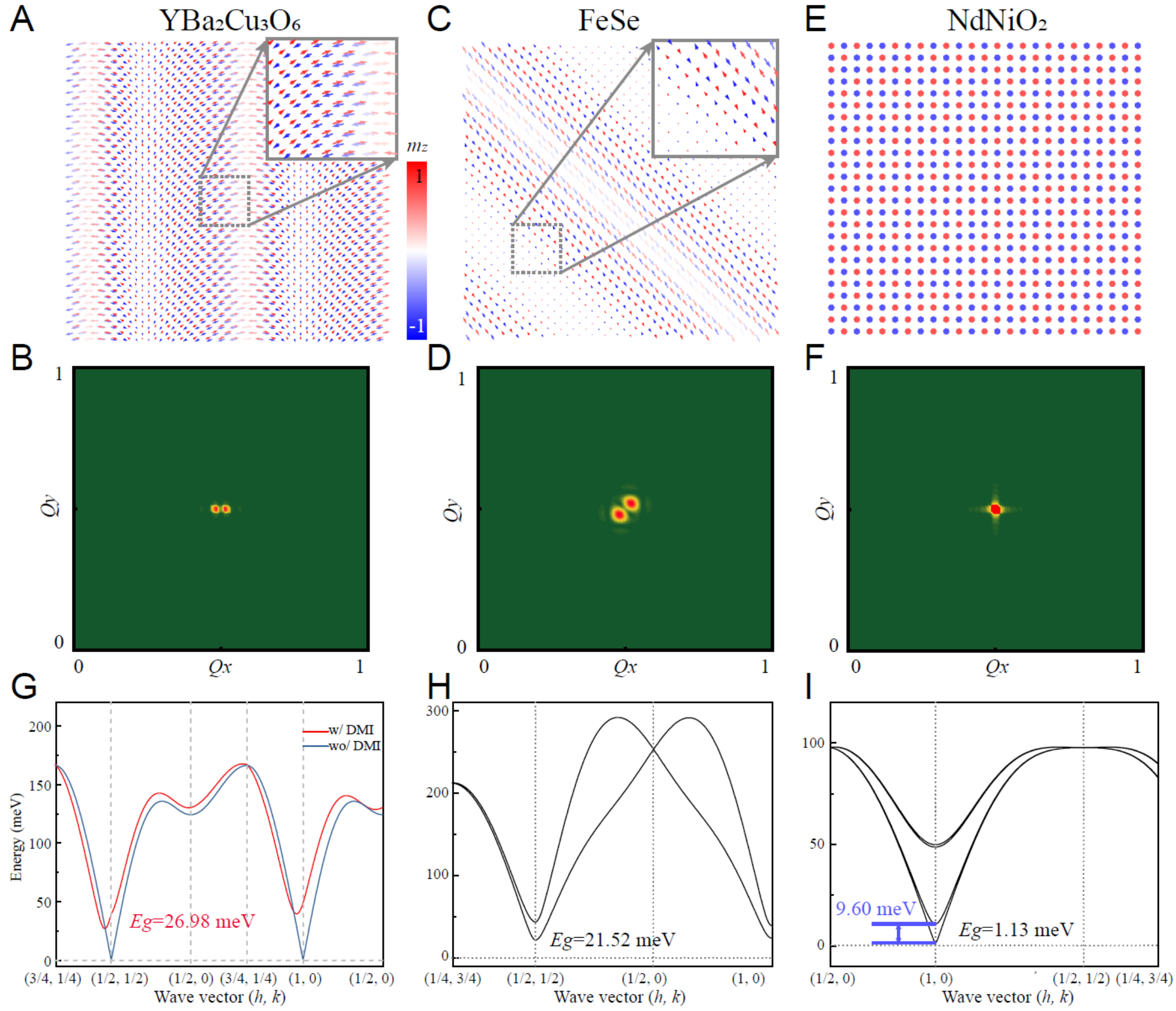

**Fig. 2. Spin textures and excitations from DMI.** Real-space spin configurations (a,c,e), reciprocal-space spin fluctuations (b,d,f), and spin wave spectra (g,h,i) for $YBa_2Cu_3O_6$, FeSe, and $NdNiO_2$. Incommensurate order, spin wave gaps, and noncollinear textures are captured by the Hamiltonian including DMI (Equation 1).

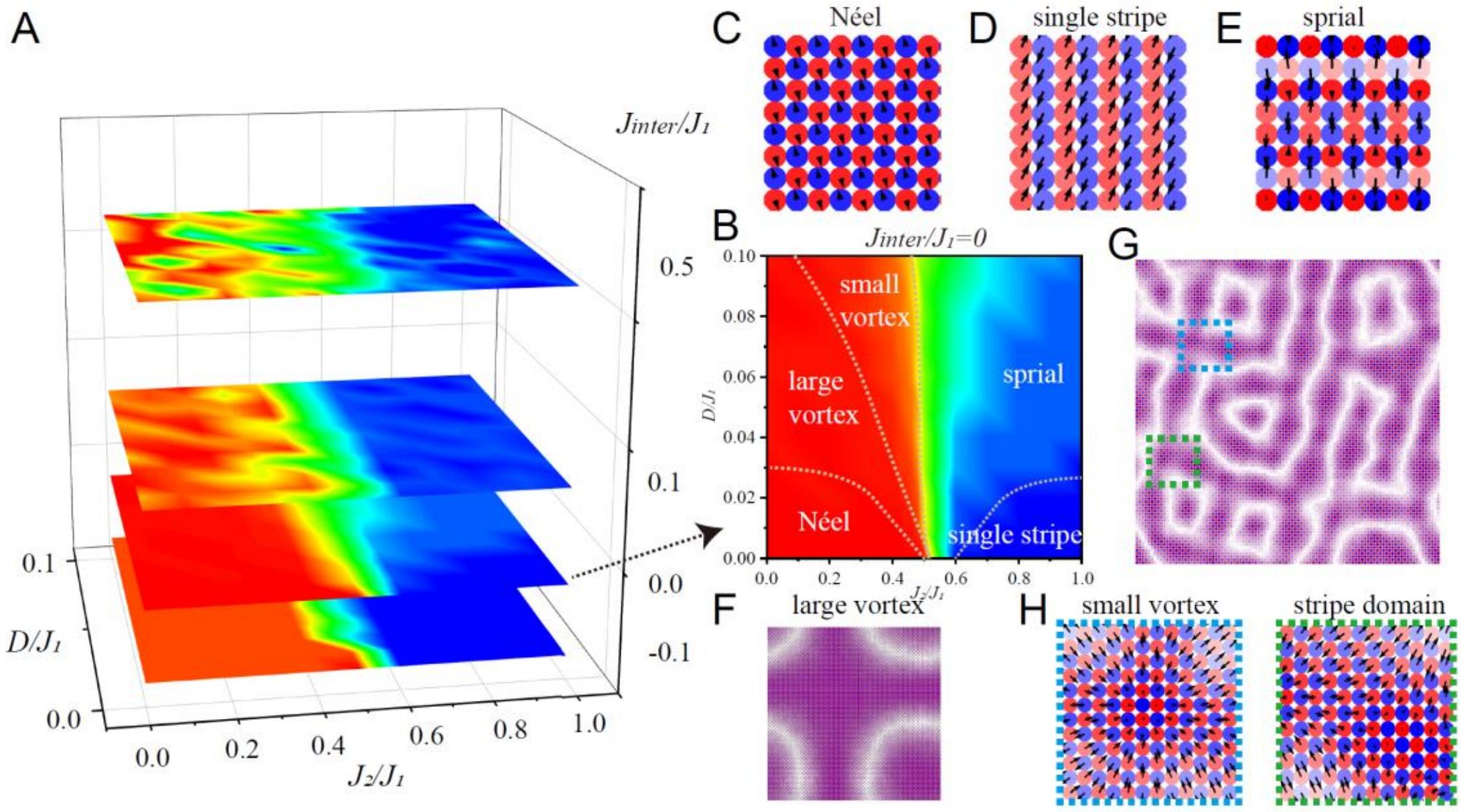

**Fig. 3. Magnetic phase diagram from MC simulations including DMI. a**, Phase diagram of spin textures in the $J_1$–$J_2$–$J_{\text{inter}}$–$D$ parameter space. Colors represent the peak intensity of the Fourier-transformed reciprocal wave vector $Q^2$ in the square lattice. **b**, Phase diagram for $J_{\text{inter}}/J_1 = 0$, showing four distinct magnetic states. Real-space spin configurations for **c**, Néel order, **d**, single-stripe order, **e**, spiral order. **f**, large vortex, and **g**, small vortex. **h**, Zoomed-in view of a single vortex and stripe domain in (g). Red-blue color scale and arrows denote out-of-plane and in-plane spin components, respectively. The emergence of vortex states between Néel and stripe phases highlights the crucial role of DMI in generating noncollinear spin textures.

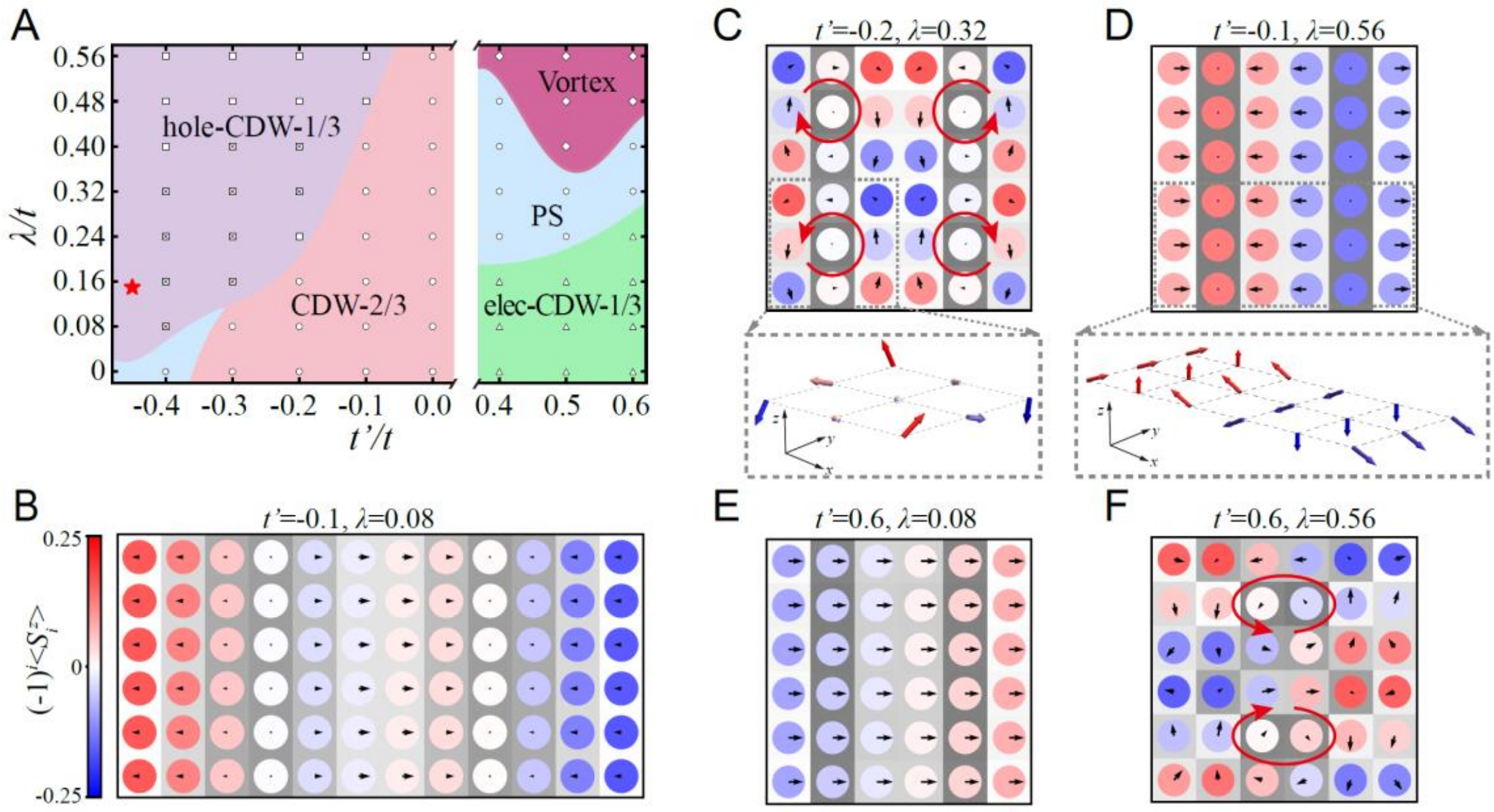

**Fig. 4. Ground state phase diagram for the Rashba–Hubbard model.** Calculations are performed on a cylindrical lattice of size $L_x = 12$, $L_y = 6$ at a doping level $\delta_h = 1/9$. **a**, Ground-state phase diagram showing five distinct phases. In the hole-CDW-$1/3$ phase, regions where vortexes form are marked with "×" symbols. **b–f**, Spin textures and charge distributions for five representative parameter sets. In panels **c–f**, results are extracted from the central $6 \times 6$ region of the square lattice for clarity. Vectors represent the in-plane spin components $(-1)^i\langle\hat{S}_i^x\rangle$ and $(-1)^i\langle\hat{S}_i^y\rangle$. Circle colors denote the out-of-plane spin magnetization $(-1)^i\langle\hat{S}_i^z\rangle$. The greyscale background indicates electron density: grey regions where holes accumulate, white regions where electrons accumulate. The lower panels of **c** and **d** provide three-dimensional visualizations of the spin orientations in the selected region, clearly illustrating a vortex structure centered on a hole (**c**) and a spiral magnetic configuration in the $xz$-plane (**d**). The DFT-fitted model parameter set for $YBa_2Cu_3O_6$, specifically $t'/t \approx -0.45$ and $\lambda/t \approx 0.15$, is highlighted in panel **a** with star.

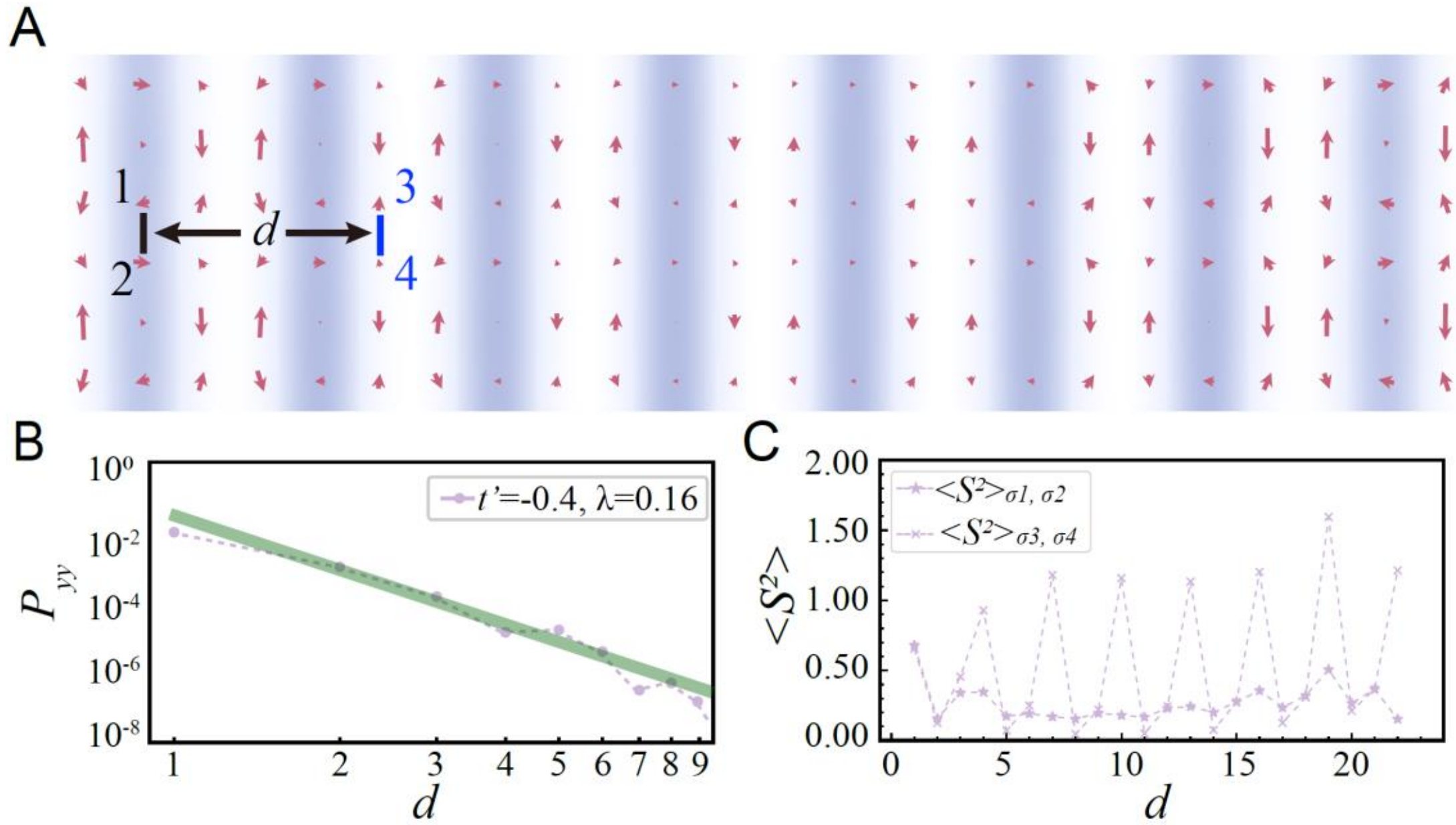

**Fig. 5. Ground state characterics of the Rashba–Hubbard model with parameters close to DFT-fitted those of $YBa_2Cu_3O_6$. a**, Spin textures in the $xy$-plane (red arrows) and grayscale hole distribution on a larger square-lattice cylinder with dimensions $L_x = 24$ and $L_y = 6$. **b**, Spin-singlet superconducting correlations as a function of distance, measured from a vertical bond (black bold line) connecting sites 1 and 2 to another vertical bond (blue bold line) connecting sites 3 and 4. The green bold line denotes the best fit to a power-law decay. **c**, Pairing angular momentum on these two bonds.

**Table I | Magnetic exchange parameters for $YBa_2Cu_3O_6$, FeSe, and $NdNiO_2$.** $J_1$, $J_2$, $J_3$, and $J_{oop}$ denote the in-plane first-, second-, third-nearest-neighbor, and out-of-plane isotropic Heisenberg exchange interactions, respectively. Positive (negative) values indicate antiferromagnetic (ferromagnetic) coupling. $D_1$ and $D_2$ are the first- and second-nearest-neighbor DMI strengths. $A$ is the single-ion anisotropy. All values are in meV.

| | $J_1$ | $J_2$ | $J_3$ | $J_{oop}$ | $D_1$ | $D_2$ | $A$ |
|---|---|---|---|---|---|---|---|
| $YBa_2Cu_3O_6$ | 104.877 | 21.489 | NA | / | 7.956 | NA | 1.453 |
| FeSe | 52.811 | 61.210 | -0.832 | / | 0.000 | 2.843 | 0.013 |
| $NdNiO_2$ | 45.455 | 0.022 | / | 6.485 | 2.680 | 0.000 | 0.735 |